\documentclass[12pt]{article}
\usepackage{amssymb, amsmath}
\usepackage{amsfonts}
\usepackage{cite}

\newcommand{\Section}[1]%
{\section{#1}\setcounter{equation}{0}%
\setcounter{theorem}{0}}
\newtheorem{theorem}{Theorem}

\def\ze{\mathbb{Z}}

{\par\noindent{\em #1:\ }}%
{~\rule{2mm}{2mm}\par\bigskip}

\begin{document}
\topskip 1cm
\begin{center}
{\large\bf Stability of Majorana Edge Zero Modes against Interactions\\}
\bigskip\bigskip
{\large Tohru Koma}\\
\bigskip\medskip
{\small Department of Physics, Gakushuin University (retired), Mejiro, Toshima-ku, Tokyo 171-8588, JAPAN\\}
\end{center}
\bigskip\bigskip\bigskip

\noindent
{\bf Abstract:} We study an interacting Majorana chain with an open boundary condition. 
In the case without interactions, the system shows a prototypical Majorana edge zero mode 
in the sector of the ground state with a spectral gap above the sector. 
We prove that both of the Majorana edge zero mode and the non-vanishing spectral gap are stable against 
generic weak interactions whose fermion parity is even. We also deal with the corresponding ladder systems, 
and discuss the $\mathbb{Z}_2$ index for the Majorana edge zero modes.

\newpage
\tableofcontents
\newpage

\Section{Introduction}

Majorana fermions have been in the focus of many researchers, so far. 
In particular, Majorana edge zero modes \cite{Kitaev} are the most typical ones in condensed matter physics. 
Although they exhibit the peculiarity, i.e., one half of a usual complex fermion,
they are indeed realized at open boundaries of tight binding models. 
However, the effects of interactions are still unclear. Fidkowski and Kitaev \cite{FK} discussed 
the effects of interactions for Majorana ladders from the point of view of topological classification. 
They concluded that the topological invariant $\ze$ for the free fermion classification is broken to $\ze_8$,   
due to the introduction of generic interactions, under a certain assumption on the symmetry of the Hamiltonian. 
Katsura, Schuricht and Takahashi \cite{KST} obtained the explicit forms of the left and right Majorana edge 
zero modes for a one-parameter family of interacting Majorana chains which are derived from 
a certain XYZ spin chain in a magnetic field, via Jordan-Wigner transformation.   
Fendley \cite{Fendley} obtained the zero mode operator for the exactly solvable XYZ chain. 
Moreover, the parafermionic edge zero modes which are a generalization of the Majorana edge zero mode 
were treated in \cite{Fendley2,Fendley3,ARFGB}. For those zero modes, 
the dynamics \cite{YAM,YAM2} and the thermal effect \cite{EFKN} have been studied as well. 
See also \cite{XZ,XZ2,XZ3,AP,CMA,CMSS,Bomantara,WHKS} for other several aspects of parafermions. 
However, mathematical treatments of Majorana edge zero modes for interacting systems are still rare. 

In the present paper, we study an interacting Majorana chain and ladders with an open boundary. 
In the case without interactions, 
these systems show Majorana edge zero modes in the sector of the ground state with a spectral gap above the sector.  
For the single chain, 
we prove that both of the Majorana edge zero mode and the non-vanishing spectral gap are stable against generic weak interactions 
that have even fermion parity. The results are summarized as Theorem~\ref{mainTheorem} in Sec.~\ref{Sec:MHchain} below. 

In Sec.~\ref{Sec:ladder}, we deal with the ladders with $L$ legs for $L\ge 2$. 
Clearly, for the $L$ copies of the noninteracting Majorana chains, there appear $L$ Majorana edge zero modes at the left edge. 
When $L$ is even, certain interchain interactions lift all the degeneracy   
for these Majorana edge zero modes. On the other hand, when $L$ is odd, there remains an unpaired Majorana edge 
zero mode for any weak interaction that has even fermion parity. Namely, the evenoddness of the Majorana edge 
zero modes is invariant against weak perturbations with even parity. This implies that for adding Majorana 
edge zero modes, the corresponding index is given by the additive group $\ze_2$ 
in the sense of the stability against generic weak perturbations.

\bigskip\bigskip

\noindent
{\bf Acknowledgements:} I would like to thank Tomonari Mizoguchi for many useful discussions. 
I also thank Hosho Katsura for many helpful comments. 


\Section{Preliminaries}
\label{Sec:Prelim}

Consider first a prototypical Majorana chain whose Hamiltonian is given by \cite{Kitaev,FK} 
\begin{equation}
\label{H0}
H_0:=i\kappa \sum_{\ell=1}^{N-1} c_{2\ell-1}c_{2\ell}+i\sum_{\ell=1}^{N-1}c_{2\ell}c_{2\ell+1}, 
\end{equation}
where $\kappa$ is a real parameter, and $c_i$ is the Majorana fermion operator at the site $i\in\{1,2,\ldots,2N\}$ 
with the length $2N$ of the chain. 
The Majorana fermion operators satisfy $c_i^\dagger=c_i$ and obey the anticommutation relations, 
$$
\{c_i,c_j\}=2\delta_{i,j}.
$$
We assume $0\le|\kappa|<1$ so that there appears a spectral gap above the ground state of $H_0$. 
Since the Hamiltonian $H_0$ does not contain $c_{2N}$, $c_{2N}$ commutes with $H_0$. 
Namely, the right edge mode is trivially given by $c_{2N}$.  
 
In order to obtain the Majorana edge zero mode $\gamma_0$ at the left edge for the Hamiltonian $H_0$, we set 
\begin{equation}
\gamma_0=\sum_{m=1}^{N}u_m c_{2m-1}.
\end{equation}
The coefficients $u_m$ are determined so as to satisfy the commutation relation $[H_0,\gamma_0]=0$.  
Clearly, when this relation is fulfilled, the state $\gamma_0|0\rangle$ has the same energy for the ground state $|0\rangle$,  
i.e., it is the Majorana zero mode. Note that
$$
[H_0,\gamma_0]=i\kappa\sum_{\ell=1}^{N-1}\sum_{m=1}^N[c_{2\ell-1}c_{2\ell},c_{2m-1}]u_m
+i\sum_{\ell=1}^{N-1}\sum_{m=1}^N[c_{2\ell}c_{2\ell+1},c_{2m-1}]u_m.
$$
The commutators in the summands are computed as follows: 
$$
c_{2\ell-1}c_{2\ell}c_{2m-1}-c_{2m-1}c_{2\ell-1}c_{2\ell}=-2\delta_{\ell,m}c_{2\ell},
$$
and 
$$
c_{2\ell}c_{2\ell+1}c_{2m-1}-c_{2m-1}c_{2\ell}c_{2\ell+1}=2\delta_{\ell+1,m}c_{2\ell}.
$$
Substituting these into the above the right-hand side, one has 
$$
[H_0,\gamma_0]=-2i\kappa\sum_{\ell=1}^{N-1}c_{2\ell}u_\ell+2i\sum_{\ell=1}^{N-1}c_{2\ell}u_{\ell+1}=0.
$$
This implies $u_{\ell+1}=\kappa u_\ell$ for $\ell=1,2,\ldots,N-1$. Therefore, the coefficients $u_\ell$ are 
given by $u_\ell=\kappa^{\ell-1}u_1$ with a constant $u_1$. As a result, one obtains 
\begin{equation}
\label{gamma0}
\gamma_0=\sum_{\ell=1}^N\kappa^{\ell-1}c_{2\ell-1},
\end{equation}
where we have set $u_1=1$ for simplicity. Since $|\kappa|<1$ from the assumption, 
the mode is localized at the left edge of the chain. The normalized operator is given by 
\begin{equation}
\label{hatgamma0}
\hat{\gamma}_0:=\sqrt{\frac{1-\kappa^2}{1-\kappa^{2N}}}\gamma_0,
\end{equation}
which satisfies $\hat{\gamma}_0^2=1$. 

Let us consider the situation when the Hamiltonian $H_0$ is diagonalized in terms of the usual complex fermions. 
The two Majorana edge modes are also treated in terms of the corresponding complex single fermion. 
For this purpose, we introduce a fermion operator, 
\begin{equation}
\label{aE}
a_{\rm E}:=\frac{1}{2}(\hat{\gamma}_0+ic_{2N}).
\end{equation}
One can check $\{a_{\rm E},a_{\rm E}^\dagger\}=1$, and $a_{\rm E}^2=0$. Let $|0\rangle$ be the ground state of $H_0$ 
which satisfies $a_{\rm E}|0\rangle=0$. Then, the other ground state is given by $a_{\rm E}^\dagger|0\rangle$. 
Clearly, one has $\langle 0|a_{\rm E}^\dagger|0\rangle=0$. 
This implies that the sector of the ground state of $H_0$ is spanned by $|0\rangle$ and $a_{\rm E}^\dagger|0\rangle$.

\Section{Majorana-Hubbard chains}
\label{Sec:MHchain}

Next we introduce interactions $V$ for the free Hamiltonian $H_0$ of (\ref{H0}). 
More specifically, we consider the Majorana-Hubbard chains \cite{RF} whose generic form of the Hamiltonians is given by 
\begin{equation}
\label{Hg}
H_g=H_0+gV
\end{equation}
with the interaction Hamiltonian,  
\begin{equation}
\label{V}
V=\sum_{i<j<k<\ell}K_{ijk\ell}c_ic_jc_kc_\ell,
\end{equation}
where $g$ and $K_{ijk\ell}$ are real parameters. For simplicity, we assume that the interaction $V$ is of finite range. 
In our approach, we can treat more general interactions 
with even parity such as many-body interactions consisting of six or eight Majorana fermion operators. 
We also assume that $V$ dose not contain $c_{2N}$. Therefore, $c_{2N}$ is still the right edge mode. 

To begin with, we present a general argument about the Majorana edge zero modes. 
When $g=0$ for the Hamiltonian $H_g=H_0+gV$, the sector of the ground state is spanned by the ground state $|0\rangle$ 
and $a_{\rm E}^\dagger|0\rangle$ as mentioned at the end of the preceding section. 
We write $P(0)$ for the projection onto the sector of the ground state of $H_0$.  

In Appendix~\ref{StabSpeGap}, we prove that the spectral gap above 
the sector of the ground state is stable against weak perturbations \cite{Hastings,RS,Koma} 
under the assumption that the sector of the ground state is two-fold degenerate. This degeneracy will be proved below 
in this section. Namely, there exists a positive constant $g_{\rm max}$ such that for $|g|\le g_{\rm max}$, 
a non-vanishing spectral gap above the sector of the ground state exists. 
Therefore, for $|g|\le g_{\rm max}$, there exists a unitary operator $U(g)$ such that 
the spectral projection $P(g)$ onto the low energy sector for the Hamiltonian $H_g$ is given by \cite{Kato,BMNS}
\begin{equation}
P(g)=U(g)P(0)U(g)^\dagger. 
\end{equation}
These low energy states are given by 
$$
|0,g\rangle:=U(g)|0\rangle \quad \mbox{and} \quad  a_{\rm E}(g)^\dagger|0,g\rangle,
$$
where have written $a_{\rm E}(g):=U(g)a_{\rm E} U(g)^\dagger$.  
In fact, from the parity conservation, these two states are an eigenstate of $H_g$ and orthogonal to each other. 
However, it is not clear whether or not the energies are degenerate.  

Let us consider the eigenvalue equation, 
\begin{equation}
\label{eigenEq}
H_g |0,g\rangle=E_0(g)|0,g\rangle, 
\end{equation}
where $E_0(g)$ is the eigenenergy of $|0,g\rangle$. 
By multiplying the both sides by $c_{2N}$ and using $[H_g,c_{2N}]=0$, one has 
$$
H_g c_{2N}|0,g\rangle=E_0(g)c_{2N}|0,g\rangle. 
$$
Therefore, the state $c_{2N}|0,g\rangle$ is an eigenstate of $H_g$ with the same eigenvalue $E_0(g)$. 
However, it has the opposite fermion parity to that of $|0,g\rangle$. 
Since such a state is unique and must be equal to $a_{\rm E}(g)^\dagger|0,g\rangle$. As a result, we have 
$$
H_g a_{\rm E}(g)^\dagger |0,g\rangle=E_0(g)a_{\rm E}(g)^\dagger |0,g\rangle.  
$$
Thus, the two states are degenerate. Further, this can be rewritten as 
$$
[H_g,a_{\rm E}(g)^\dagger]|0,g\rangle=0.
$$
The operator $a_{\rm E}(g)^\dagger$ in the commutator can be written   
$$
a_{\rm E}(g)^\dagger = U(g) a_{\rm E}^\dagger U(g)^\dagger =\frac{1}{2}U(g)(\hat{\gamma}_0-ic_{2N})U(g)^\dagger
=\frac{1}{2}U(g)\hat{\gamma}_0 U(g)^\dagger-\frac{i}{2}c_{2N}
$$
{from} the definitions and the assumption $[H_g,c_{2N}]=0$ on the Hamiltonian $H_g$.  These yield   
\begin{equation}
\label{HcommuUgammaU0}
[H_g,U(g)\hat{\gamma}_0 U(g)^\dagger]|0,g\rangle=0.
\end{equation}
Thus, the operator $U(g)\hat{\gamma}_0 U(g)^\dagger$ creates the Majorana edge zero mode above 
the ground state $|0,g\rangle$ of the Hamiltonian $H_g$. 
Namely, the Majorana edge zero mode is stable against the perturbations.

The above result (\ref{HcommuUgammaU0}) suggests that the operator relation $[H_g,U(g)\hat{\gamma}_0 U(g)^\dagger]=0$ holds 
without acting on the state $|0,g\rangle$. 
In fact, Goldstein and Chamon \cite{GC} proved that there exists a non-trivial operator $\gamma$ with odd fermion parity  
such that it satisfies $[H,\gamma]=0$ for a generic Hamiltonian $H$ 
which consists of an odd number of Majorana fermion operators. 
If the operator $\gamma$ satisfies $[H,\gamma]=0$, then $[H,\gamma^\dagger]=0$. This yields $[H,\gamma +\gamma^\dagger]=0$. 
Therefore, one can take $\gamma$ to satisfy $\gamma^\dagger=\gamma$. The proof of the existence of $\gamma$ 
is given in Appendix~\ref{existencegamma}. 

Since the present Hamiltonian $H_g$ consists of the odd number of the Majorana fermions, 
there exists an operator $\hat{\gamma}_g$ such that $[H_g,\hat{\gamma}_g]=0$. 
Clearly, this operator $\hat{\gamma}_g$ also creates the Majorana edge zero mode at the left edge.  
Therefore, we have 
\begin{equation}
U(g)\hat{\gamma}_0 U(g)^\dagger|0,g\rangle=\hat{\gamma}_g|0,g\rangle 
\end{equation}
for a small coupling constant $g$. The locality of the Majorana edge mode is proved in Appendix~\ref{LocalMajoranaedge}. 
Thus, our main results are summarized as follows: 

\begin{theorem}
\label{mainTheorem}
There exists a positive constant $g_{\rm max}$ such that for $|g|\le g_{\rm max}$, the ground state of the Hamiltonian 
$H_g$ of (\ref{Hg}) is two-fold degenerate with a non-vanishing spectral gap above the sector of the ground state. 
One of the two ground-state vectors can be taken to be a Majorana edge mode which is localized near the left edge. 
\end{theorem}

\Section{Majorana-Hubbard ladders and edge $\ze_2$ index}
\label{Sec:ladder}

In this section, we deal with Majorana ladders \cite{FK} which are a system of $L$ Majorana chains with 
interchain interactions. Here, $L\ge 2$ is the number of the legs of the ladder. 
As we will show below, we can expect that the evenoddness of the Majorana left edge modes is stable 
against generic weak perturbations with even fermion parity. 

Consider first the case with $L=2$. The Hamiltonian is given by 
\begin{equation}
\label{H0L2}
H_0^{(2)}:=i \sum_{j=1}^2\sum_{\ell=1}^{N-1} (\kappa c_{2\ell-1,j}c_{2\ell,j}+c_{2\ell,j}c_{2\ell+1,j}), 
\end{equation}
which is the sum of the two copies of the Majorana chain of (\ref{H0}). 
The two copies of the set of the Majorana operators satisfy the anti-commutation relations, 
\begin{equation}
\{c_{m,i},c_{n,j}\}=2\delta_{m,n}\delta_{i,j},\quad \mbox{for } m,n=1,2,\ldots,2N-1; \ i,j=1,2. 
\end{equation}
The two operators, $c_{2N,j}$ $j=1,2$, are the trivial right edge modes.  
Clearly, the two left Majorana edge modes are given by 
\begin{equation}
\label{gamma0j}
\gamma_{0,j}=\sum_{\ell=1}^N\kappa^{\ell-1}c_{2\ell-1,j}\quad \mbox{for } j=1,2. 
\end{equation}
The normalized operators are given by 
\begin{equation}
\label{hatgamma0j}
\hat{\gamma}_{0,j}:=\sqrt{\frac{1-\kappa^2}{1-\kappa^{2N}}}\gamma_{0,j},
\end{equation}
for $j=1,2$, which satisfy $(\hat{\gamma}_{0,j})^2=1$ for $j=1,2$. 

Let us consider a very simple interchain interaction, 
\begin{equation}
V=i\hat{\gamma}_{0,1}\hat{\gamma}_{0,2},  
\end{equation}
which is instructive to understand the difference between the ladder with two legs and the single chain. 
Clearly, this interaction has even fermion parity and is of short range. 

In order to diagonalize the total Hamiltonian $H_g=H_0^{(2)}+gV$ in terms of usual complex fermions, 
we introduce a complex fermion operator, 
\begin{equation}
\chi:=\frac{1}{2}(\hat{\gamma}_{0,1}+i\hat{\gamma}_{0,2}).
\end{equation}
Then, one has 
\begin{equation}
\{\chi,\chi^\dagger\}=1
\end{equation}
and 
\begin{equation}
i\hat{\gamma}_{0,1}\hat{\gamma}_{0,2}=2\chi^\dagger\chi-1. 
\end{equation}
This implies that the two eigenvalues of $V$ are given by $\pm 1$. Similarly, for the right edge modes, 
we can introduce the corresponding complex fermion $(c_{2N,1}+i c_{2N,2})/2$. 
Therefore, the degeneracy of the left zero modes is lifted by the perturbation $V$ even for any small coupling $g$. 

Clearly, this result can be extended to the generic ladders with even number legs.  
On the other hand, for odd number legs, there must remain a unpaired Majorana fermion. 
Thus, the evenoddness of Majorana edge modes can be expected to be the $\ze_2$ invariant 
under weak perturbations with even parity.

\Section{Explicit construction of Majorana edge zero modes}

\subsection{Exactly solvable cases}

Clearly, from the result $[H_0,\gamma_0]=0$ of Sec.~\ref{Sec:Prelim}, 
one has $[H_g,\gamma_0]=[H_0,\gamma_0]+g[V,\gamma_0]=g[V,\gamma_0]$. Therefore, if $[V,\gamma_0]=0$, 
then $\gamma_0$ is the desired solution. 

Since the operator $\gamma_0$ consists of $c_i$ at only the odd site $i=2\ell-1$, 
$\gamma_0$ commutes with an interaction $V$ which consists of $c_j$ at only the even site $j=2k$. Namely, 
the interaction $V$ has the form,  
$$
V=\sum_{i<j<k<\ell}K_{2i,2j,2k,2\ell}c_{2i}c_{2j}c_{2k}c_{2\ell}. 
$$
In a sense, it may be said that this is a trivial case. 

In order to present other examples, we introduce an operator, 
$$
b_\ell:=\kappa c_{2\ell-1}-c_{2\ell+1}. 
$$
Note that 
\begin{eqnarray*}
& &\{b_\ell,c_{2\ell-1}+\kappa c_{2\ell+1}\}\\
&=&(\kappa c_{2\ell-1}-c_{2\ell+1})(c_{2\ell-1}+\kappa c_{2\ell+1})
+(c_{2\ell-1}+\kappa c_{2\ell+1})(\kappa c_{2\ell-1}-c_{2\ell+1})\\
&=&\kappa+\kappa^2c_{2\ell-1}c_{2\ell+1}-c_{2\ell+1}c_{2\ell-1}-\kappa +\kappa-c_{2\ell-1}c_{2\ell+1}
+\kappa^2c_{2\ell+1}c_{2\ell-1}-\kappa\\
&=&0. 
\end{eqnarray*}
Using this identity, we have 
\begin{eqnarray*}
\{b_\ell,\gamma_0\}&=&\sum_{m=1}\{b_\ell,\kappa^{m-1}c_{2m-1}\}\\
&=&\{b_\ell,\kappa^{\ell-1}c_{2\ell-1}+\kappa^\ell c_{2\ell+1}\}\\
&=&\kappa^{\ell-1}\{b_\ell,c_{2\ell-1}+\kappa c_{2\ell+1}\}=0.
\end{eqnarray*}
As an interaction $V$, let us consider 
$$
V=\sum_{\ell=1} b_\ell c_{2\ell}c_{2\ell+2}c_{2\ell+4}
=\sum_{\ell=1}(\kappa c_{2\ell-1}-c_{2\ell+1})c_{2\ell}c_{2\ell+2}c_{2\ell+4}.
$$
{From} the above observations, we have $[\gamma_0,V]=0$. 

Similarly, we can consider 
$$
V=\sum_{\ell=1} [b_\ell b_{\ell+1}c_{2\ell}c_{2\ell+2}+\mbox{h.c.}].
$$
This also satisfies $[\gamma_0,V]=0$. In passing, we note that 
\begin{eqnarray*}
b_\ell b_{\ell+1}c_{2\ell}c_{2\ell+2}+\mbox{h.c.}&=&b_\ell b_{\ell+1}c_{2\ell}c_{2\ell+2}+c_{2\ell+2}c_{2\ell}b_{\ell+1}b_\ell\\
&=&(b_\ell b_{\ell+1}-b_{\ell+1}b_\ell)c_{2\ell}c_{2\ell+2}\\
&=&2(\kappa^2c_{2\ell-1}c_{2\ell+1}-\kappa c_{2\ell-1}c_{2\ell+3}+c_{2\ell+1}c_{2\ell+3})c_{2\ell}c_{2\ell+2},
\end{eqnarray*}
where we have used 
\begin{eqnarray*}
b_\ell b_{\ell+1}-b_{\ell+1}b_\ell
&=&(\kappa c_{2\ell-1}-c_{2\ell+1})(\kappa c_{2\ell+1}-c_{2\ell+3})-(\kappa c_{2\ell+1}-c_{2\ell+3})(\kappa c_{2\ell-1}-c_{2\ell+1})\\
&=&2\kappa^2c_{2\ell-1}c_{2\ell+1}-2\kappa c_{2\ell-1}c_{2\ell+3}+2c_{2\ell+1}c_{2\ell+3}.
\end{eqnarray*}

\subsection{A series expansion method}

When the coupling constant $g$ of the interaction is weak compared to the spectral gap above the ground-state sector 
of the unperturbed Hamiltonian $H_0$, a series expansion method can be expected to work well \cite{RS,GMP}. 

In order to find $\gamma$ satisfying $[H_0+gV,\gamma]=0$ for a given interaction $V$ with a weak coupling $g$, 
we expand $\gamma$ into power series of $g$ as follows: 
$$
\gamma=\sum_{n=0} g^n \gamma_n.
$$
Substituting this into $[H_0,\gamma]=-g[V,\gamma]$, one has 
$$
\sum_{n=0}g^n[H_0,\gamma_n]=-\sum_{n=0}g^{n+1}[V,\gamma_n].
$$
The zeroth order is given by 
$$
[H_0,\gamma_0]=0.
$$
This is already fulfilled. The $n$-th order is given by 
\begin{equation}
\label{n-commu}
[H_0,\gamma_n]=-[V,\gamma_{n-1}]
\end{equation}
for $n=1,2,\ldots$.

As a first demonstration, we consider a very simple case $V=c_1c_2c_3c_4$. 
It consists of only a single term, but it can be expected to represent one of essential, effective interactions 
because the Majorana edge mode is localized at the edge.  

Let us consider the right-hand side of (\ref{n-commu}) in the case of $n=1$. 
Note that 
\begin{eqnarray*}
[V,\gamma_0]&=&[c_1c_2c_3c_4,c_1]+\kappa[c_1c_2c_3c_4,c_3]\\
&=&-2c_2c_3c_4-2\kappa c_1c_2c_4. 
\end{eqnarray*}
Therefore, in order to find $\gamma_1$, it is sufficient to solve 
\begin{equation}
\label{H0gamma1commu}
[H_0,\gamma_1]=2c_2c_3c_4+2\kappa c_1c_2c_4
\end{equation}
{from} (\ref{n-commu}). Before searching for a solution $\gamma_1$, we remark that 
there are many solutions even restricting $\gamma_1$ to a sum of terms of three Majorana fermions. 
In fact, the quantity $(\gamma_0H_0+H_0\gamma_0)/2$ commutes with $H_0$, and is a sum of terms of three Majorana 
fermions. 

Since $\gamma_0$ has the form of the power series of $\kappa$, we similarly set
\begin{equation}
\gamma_1=\sum_{j=0}\kappa^j \gamma_1^{(j)}. 
\end{equation}
Note that 
\begin{eqnarray}
\label{H0gamma1expand}
[H_0,\gamma_1]&=&\kappa[H_{0,1},\gamma_1]+[H_{0,0},\gamma_1]\nonumber\\
&=&\sum_{j=0}\kappa^{j+1}[H_{0,1},\gamma_{1}^{(j)}]+\sum_{j=0}\kappa^j[H_{0,0},\gamma_{1}^{(j)}],
\end{eqnarray}
where we have written 
$$
H_0=\kappa H_{0,1}+H_{0,0}
$$
with 
$$
H_{0,1}:=i\sum_{\ell=1}c_{2\ell-1}c_{2\ell}
$$
and 
$$
H_{0,0}:=i\sum_{\ell=1}c_{2\ell}c_{2\ell+1}.
$$
Therefore, from (\ref{H0gamma1commu}) and (\ref{H0gamma1expand}), we have 
\begin{equation}
\label{H00gamma10commu}
[H_{0,0},\gamma_{1}^{(0)}]=2c_2c_3c_4,
\end{equation}
\begin{equation}
\label{1stordergammaeq}
[H_{0,1},\gamma_{1}^{(0)}]+[H_{0,0},\gamma_{1}^{(1)}]=2c_1c_2c_4,
\end{equation}
and 
\begin{equation}
\label{j+1thordergammaeq}
[H_{0,1},\gamma_{1}^{(j)}]+[H_{0,0},\gamma_{1}^{(j+1)}]=0 \mbox{\ \ for \ } j=1,2,\ldots.
\end{equation}

In order to find the solution $\gamma_{1}^{(0)}$ of (\ref{H00gamma10commu}), we try $\gamma_{1}^{(0)}=-ic_2c_3c_5$ 
as a candidate. 
Actually, we have 
\begin{eqnarray*}
[H_{0,0},\gamma_{1}^{(0)}]&=&\sum_{\ell=1}[c_{2\ell}c_{2\ell+1},c_2c_3c_5]\\
&=&[c_2c_3,c_2c_3c_5]+[c_4c_5,c_2c_3c_5]\\
&=&2c_2c_3c_4.
\end{eqnarray*}
Thus, $\gamma_{1}^{(0)}=-ic_2c_3c_5$ is one of the solutions of (\ref{H00gamma10commu}). 

Next, in order to obtain $\gamma_{1}^{(1)}$, let us compute the first commutator $[H_{0,1},\gamma_{1}^{(0)}]$ 
in the left-hand side of (\ref{1stordergammaeq}). Note that 
\begin{eqnarray*}
[H_{0,1},\gamma_{1}^{(0)}]&=&\sum_{\ell=1}[c_{2\ell-1}c_{2\ell},c_2c_3c_5]\\
&=&[c_1c_2,c_2c_3c_5]+[c_3c_4,c_2c_3c_5]+[c_5c_6,c_2c_3c_5]\\
&=&2c_1c_3c_5-2c_2c_4c_5-2c_2c_3c_6. 
\end{eqnarray*}
Substituting this into the left-hand side of (\ref{1stordergammaeq}), one has 
\begin{equation}
[H_{0,0},\gamma_{1}^{(1)}]=2c_1c_2c_4-2c_1c_3c_5+2c_2c_4c_5+2c_2c_3c_6.
\end{equation}
We prepare the following: 
$$
[H_{0,0},(-i)c_3c_4c_5]=\sum_{\ell=1}[c_{2\ell}c_{2\ell+1},c_3c_4c_5]=[c_2c_3,c_3c_4c_5]+[c_4c_5,c_3c_4c_5]
=2c_2c_4c_5,
$$
$$
[H_{0,0},(-i)c_2c_3c_7]=\sum_{\ell=1}[c_{2\ell}c_{2\ell+1},c_2c_3c_7]=[c_2c_3,c_2c_3c_7]+[c_6c_7,c_2c_3c_7]
=2c_2c_3c_6,
$$
$$
[H_{0,0},(-i)c_1c_3c_4]=\sum_{\ell=1}[c_{2\ell}c_{2\ell+1},c_1c_3c_4]=[c_2c_3,c_1c_3c_4]+[c_4c_5,c_1c_3c_4]
=2c_1c_2c_4-2c_1c_3c_5,
$$
and 
$$
[H_{0,0},(-i)c_1c_2c_5]=\sum_{\ell=1}[c_{2\ell}c_{2\ell+1},c_1c_2c_5]=[c_2c_3,c_1c_2c_5]
+[c_4c_5,c_1c_2c_5]=-2c_1c_3c_5+2c_1c_2c_4. 
$$
Therefore, we have a solution, 
$$
\gamma_{1}^{(1)}=-ic_3c_4c_5-ic_2c_3c_7-i\lambda_{1}^{(1)}c_1c_3c_4-i(1-\lambda_{1}^{(1)})c_1c_2c_5,
$$
where $\lambda_{1}^{(1)}$ is a real parameter. 

Further, $\gamma_{1}^{(2)}$ is determined by 
\begin{equation}
\label{gamma12eq}
[H_{0,0},\gamma_{1}^{(2)}]=-[H_{0,1},\gamma_{1}^{(1)}]
\end{equation}
{from} the equation (\ref{j+1thordergammaeq}) for $j=1$. The right-hand side is calculated as 
\begin{eqnarray}
\label{commuH00gamma11}
-[H_{0,1},\gamma_{1}^{(1)}]&=&-\sum_{\ell=1}[c_{2\ell-1}c_{2\ell},c_3c_4c_5+c_2c_3c_7+\lambda_{1}^{(1)}c_1c_3c_4+(1-\lambda_{1}^{(1)})
c_1c_2c_5]\nonumber\\
&=&2c_3c_4c_6-2c_1c_3c_7+2c_2c_4c_7+2c_2c_3c_8+2\lambda_{1}^{(1)}c_2c_3c_4+2(1-\lambda_{1}^{(1)})c_1c_2c_6.\nonumber\\
\end{eqnarray}
Note that 
\begin{equation}
\label{commuH00c239}
[H_{0,0},(-i)c_2c_3c_9]=\sum_{\ell=1}[c_{2\ell}c_{2\ell+1},c_2c_3c_9]=2c_2c_3c_8,
\end{equation}
and
\begin{equation} 
\label{commuH00c127}
[H_{0,0},(-i)c_1c_2c_7]=[c_2c_3,c_1c_2c_7]+[c_6c_7,c_1c_2c_7]=-2c_1c_3c_7+2c_1c_2c_6.
\end{equation}
By choosing $\lambda_{1}^{(1)}=0$, we can obtain the three terms in the right-hand side of (\ref{commuH00gamma11}), 
and the rest are $2c_3c_4c_6$ and $2c_2c_4c_7$. 

In order to obtain the corresponding contributions to the rest of the two terms, we prepare the following: 
$$
I_{246}:=[H_{0,0},(-i)c_2c_4c_6]=-2c_3c_4c_6-2c_2c_5c_6-2c_2c_4c_7,
$$
$$
I_{257}:=[H_{0,0},(-i)c_2c_5c_7]=-2c_3c_5c_7+2c_2c_5c_6+2c_2c_4c_7,
$$
$$
I_{347}:=[H_{0,0},(-i)c_3c_4c_7]=2c_3c_4c_6-2c_3c_5c_7+2c_2c_4c_7,
$$
and
$$
I_{356}:=[H_{0,0},(-i)c_3c_5c_6]=2c_3c_4c_6-2c_3c_5c_7+2c_2c_5c_6.
$$
{From} these results, one has 
$$
2I_{246}+I_{257}-2I_{347}+I_{356}=-6c_3c_4c_6-6c_2c_4c_7.
$$
This implies 
$$
[H_{0,0},(\frac{2}{3}ic_2c_4c_6+\frac{1}{3}ic_2c_5c_7-\frac{2}{3}ic_3c_4c_7+\frac{1}{3}ic_3c_5c_6)]
=2c_3c_4c_6+2c_2c_4c_7.
$$
Combining this, (\ref{gamma12eq}), (\ref{commuH00gamma11}), (\ref{commuH00c239}) and (\ref{commuH00c127}), we obtain  
$$
\gamma_{1}^{(2)}=\frac{2}{3}ic_2c_4c_6+\frac{1}{3}ic_2c_5c_7-\frac{2}{3}ic_3c_4c_7+\frac{1}{3}ic_3c_5c_6
-ic_1c_2c_7-ic_2c_3c_9. 
$$
Thus, we can obtain the approximate solution of the edge mode by the series expansion method. 

\appendix 

\Section{Stability of the spectral gap}
\label{StabSpeGap}

In this appendix, we prove that the gap above the two-fold degenerate ground state of the unperturbed 
Hamiltonian $H_0$ does not close under weak interactions. As proved in Sec.~\ref{Sec:MHchain}, 
if there exists a non-vanishing spectral gap above the sector of the ground state which are spanned by  
the ground-state vector and the Majorana edge mode, then these two energies are always degenerate. 
We assume that for a sufficiently small $|g|$, 
there exists a spectral gap $\Delta\tilde{E}>0$ above the two-fold degenerate ground state which continuously 
connects with the spectral gap in the case of $g=0$. Under this assumption, we will estimate $|g|$ and $\Delta\tilde{E}$.   

In order to separate the edge modes $\gamma_0$ and $c_{2N}$ from the bulk variables 
so that the two edge variables are independent of the bulk variables, 
we introduce new variables, $\tilde{c}_3,\tilde{c_5},\ldots,\tilde{c}_{2N-1}$: 
\begin{equation}
\label{tildec2ell+1}
\mathcal{N}_{2\ell+1}\tilde{c}_{2\ell+1}=\sum_{m=1}^\ell \kappa^{m-1}c_{2m-1}
-\frac{\sum_{m=1}^\ell \kappa^{2(m-1)}}{\kappa^\ell}c_{2\ell+1}
\end{equation}
for $\ell=1,2,\ldots,N-1$, where $\mathcal{N}_{2\ell+1}$ is the normalization constant 
so that $\tilde{c}_{2\ell+1}$ satisfies $(\tilde{c}_{2\ell+1})^2=1$. 
We recall the expression of the left edge mode, 
\begin{equation}
\label{gamma0recall}
\gamma_0=\sum_{\ell=1}^N \kappa^{\ell-1}c_{2\ell-1}. 
\end{equation}
{From} these expressions, one has 
\begin{equation}
\{\gamma_0,\tilde{c}_{2\ell+1}\}=0
\end{equation}
for $\ell=1,2,\ldots,N-1$, and 
\begin{equation}
\{\tilde{c}_{2\ell+1},\tilde{c}_{2\ell'+1}\}=0\quad \mbox{for \ \ } \ell\ne \ell'.
\end{equation}
Since $\mathcal{N}_{2\ell+1}\sim \kappa^{-\ell}$ for a small $\kappa$, the operator $\tilde{c}_{2\ell+1}$ is still 
localized around the site $2\ell+1$. 

To rewrite the Hamiltonian $H_0$ in terms of the variables $\tilde{c}_{2\ell+1}$, we note that 
\begin{equation}
\label{hamrewrite}
H_0=i\kappa \sum_{\ell=1}^{N-1} c_{2\ell-1}c_{2\ell}+i\sum_{\ell=1}^{N-1}c_{2\ell}c_{2\ell+1}
=i\sum_{\ell=1}^{N-1}(\kappa c_{2\ell-1}-c_{2\ell+1})c_{2\ell}. 
\end{equation}
{From} (\ref{tildec2ell+1}), one has 
\begin{equation}
\mathcal{N}_{2\ell+1}\tilde{c}_{2\ell+1}=\mathcal{N}_{2\ell-1}\tilde{c}_{2\ell-1}
+\frac{ \sum_{m=1}^\ell \kappa^{2(m-1)} }{\kappa^\ell} (\kappa c_{2\ell-1}-c_{2\ell+1})
\end{equation}
for $\ell=2,3,\ldots,N-1$. This yields 
\begin{equation}
\kappa c_{2\ell-1} - c_{2\ell+1}=\frac{\kappa^\ell}{\sum_{m=1}^\ell \kappa^{2(m-1)}}
(\mathcal{N}_{2\ell+1}\tilde{c}_{2\ell+1}-\mathcal{N}_{2\ell-1}\tilde{c}_{2\ell-1})
\end{equation}
for $\ell=2,3,\ldots,N-1$. Substituting this and (\ref{tildec2ell+1}) for $\ell=1$ into 
the above expression (\ref{hamrewrite}) of the Hamiltonian $H_0$, we have 
\begin{equation}
H_0=i\kappa \mathcal{N}_3\tilde{c}_3 \tilde{c}_2+ i\sum_{\ell=2}^{N-1}\frac{\kappa^\ell}{\sum_{m=1}^\ell \kappa^{2(m-1)}}
(\mathcal{N}_{2\ell+1}\tilde{c}_{2\ell+1}-\mathcal{N}_{2\ell-1}\tilde{c}_{2\ell-1})\tilde{c}_{2\ell}, 
\end{equation}
where we have written $\tilde{c}_{2\ell}=c_{2\ell}$ for $\ell=1,2,\ldots,N-1$. 
Thus, the Hamiltonian $H_0$ can be written in terms of the variables, $\tilde{c}_2,\tilde{c}_3,\ldots,\tilde{c}_{2N-1}$. 
Although the new set of the variables is given by the $2N$ variables, 
$\{\hat{\gamma}_0,\tilde{c}_2,\tilde{c}_3,\ldots,\tilde{c}_{2N-1},c_{2N}\}$, 
the expression of the Hamiltonian $H_0$ is independent of the Majorana edge zero modes, 
$\hat{\gamma}_0$ and $c_{2N}$. Namely, the two sets of the variables are separated in the expression of $H_0$. 
 
By using a pure imaginary, antisymmetric matrix $\tilde{A}$, the Hamiltonian $H_0$ can be written in the form, 
\begin{equation}
\label{H0ctilde}
H_0=\sum_{i,j\in\{2,3,\ldots,2N-1\}} \tilde{c}_i\tilde{A}_{i,j}\tilde{c}_j.
\end{equation}
By the construction, the Hamiltonian $H_0$ has the unique ground state with the non-vanishing spectral gap 
above it, except for the Majorana edge zero modes. In the following, we will treat these two ground states separately.

Following Hastings \cite{Hastings}, we introduce two copies of 
the Majorana fermions, and two copies of the Hamiltonian ${H}_0$ with opposite signs 
so that the total Hamiltonian is given by 
\begin{equation}
\label{tildeH0p}
\tilde{H}_0:=\sum_{i,j\in\{2,3,\ldots,2N-1\}} \tilde{c}_i\tilde{A}_{i,j}\tilde{c}_j
-\sum_{i,j\in\{2,3,\ldots,2N-1\}} \tilde{d}_i\tilde{A}_{i,j}\tilde{d}_j,
\end{equation}
where the Majorana operator $\tilde{d}_j$ is a copy of $\tilde{c}_j$.   
This Hamiltonian has also the unique ground state with the non-vanishing spectral gap above it. 
We write 
$$
\tilde{H}_0=
\begin{pmatrix}
\tilde{c}, & \tilde{d}
\end{pmatrix}
\begin{pmatrix}
\tilde{A} & 0 \\
0 & -\tilde{A}
\end{pmatrix}
\begin{pmatrix}
\tilde{c} \\
\tilde{d}
\end{pmatrix},
$$
and $P_\pm(\tilde{A})$ for the projection onto the positive and negative energies for $\tilde{A}$, respectively. 
We also introduce \cite{Hastings} 
$$
\hat{U}:=\frac{1}{\sqrt{2}}
\begin{pmatrix}
1 & is(\tilde{A})\\
is(\tilde{A}) & 1
\end{pmatrix},
$$
where $s(\tilde{A}):=P_+(\tilde{A})-P_-(\tilde{A})$. Then, one has 
$$
\hat{U}^\dagger
\begin{pmatrix}
\tilde{A} & 0 \\
0 & -\tilde{A}
\end{pmatrix}
\hat{U}=
\begin{pmatrix}
0 & i|\tilde{A}| \\
-i|\tilde{A}| & 0
\end{pmatrix}.
$$
We define a pair of Majorana fermions, 
\begin{equation}
\label{etaRI}
\begin{pmatrix}
\eta^{\rm R} \\
\eta^{\rm I}
\end{pmatrix}
:=\hat{U}^\dagger
\begin{pmatrix}
\tilde{c} \\
\tilde{d}
\end{pmatrix}. 
\end{equation}
{From} the above observation, one has 
$$
\tilde{H}_0=
\begin{pmatrix}
\eta^{\rm R}, & \eta^{\rm I} 
\end{pmatrix}
\begin{pmatrix}
0 & i|\tilde{A}| \\
-i|\tilde{A}| & 0
\end{pmatrix}
\begin{pmatrix}
\eta^{\rm R}\\
\eta^{\rm I}
\end{pmatrix}
=2i\eta^{\rm R}|\tilde{A}|\eta^{\rm I}.
$$
Further, we define a complex fermion, 
\begin{equation}
\label{eta}
\eta_i:=\frac{1}{2}(\eta_i^{\rm R}+i\eta_i^{\rm I}),
\end{equation}
for $i=2,3,\ldots,2N-1$. Then, 
\begin{equation}
\label{tildeH0}
\tilde{H}_0=2\eta^\dagger |\tilde{A}|\eta, 
\end{equation}
where we have dropped the constant in the right-hand side. 
Since there is a spectral gap at the zero energy in the spectrum of the matrix $\tilde{A}$, we have 
\begin{equation}
\label{lowerboundH0}
\tilde{H}_0=2\eta^\dagger |\tilde{A}|\eta\ge \Delta E_0 \eta^\dagger \eta  
\end{equation}
with a constant $\Delta E_0>0$. 

Next, consider the interaction term $V$ of the Hamiltonian $H_g$ of (\ref{Hg}). 
The interaction $V$ is written in terms of the operators, $c_1,c_2,\ldots,c_{2N-1}$. 
Our aim is to express the interaction $V$ in terms of the operators, $\eta$ and $a_{\rm E}$. 

For this purpose, we first want to express the operators, $c_1,c_3,\ldots,c_{2N-1}$, in terms of 
$\gamma_0, \tilde{c}_3,\tilde{c}_5,\ldots,\tilde{c}_{2N-1}$. 
{From} the expression (\ref{gamma0recall}) of $\gamma_0$, one has 
\begin{equation}
\label{c1}
c_1=\gamma_0-(\kappa c_3 +\kappa^2 c_5 + \kappa^3 c_7 + \cdots).
\end{equation}
Combining this with (\ref{tildec2ell+1}) for $\ell=1$, one obtains 
\begin{eqnarray}
c_3&=&\kappa c_1 -\kappa \mathcal{N}_3\tilde{c}_3\nonumber\\
&=&\kappa\gamma_0-\kappa(\kappa c_3 +\kappa^2 c_5 + \kappa^3 c_7 + \cdots)-\kappa \mathcal{N}_3\tilde{c}_3\nonumber\\
&=&\kappa\gamma_0-\kappa^2 c_3-\kappa(\kappa^2 c_5 + \kappa^3 c_7 + \cdots)-\kappa \mathcal{N}_3\tilde{c}_3. 
\end{eqnarray}
Therefore, 
\begin{equation}
\label{c3}
c_3=\frac{\kappa}{1+\kappa^2}[\gamma_0-(\kappa^2 c_5 + \kappa^3 c_7 + \cdots)]
-\frac{\kappa}{1+\kappa^2} \mathcal{N}_3\tilde{c}_3.
\end{equation}
By using this expression, we can eliminate $c_3$ from the right-hand side of (\ref{c1}). 
Similarly, from (\ref{tildec2ell+1}) for $\ell=2$, one has
\begin{eqnarray*}
\mathcal{N}_5\tilde{c}_5&=&c_1+\kappa c_3 -\frac{1+\kappa^2}{\kappa^2}c_5\\
&=&\gamma_0 -\kappa^2 c_5 -(\kappa^3 c_7 +\kappa^4 c_9 +\cdots) -\frac{1+\kappa^2}{\kappa^2}c_5.
\end{eqnarray*}
Therefore, 
\begin{equation}
c_5=\frac{\kappa^2}{1+\kappa^2+\kappa^4}[\gamma_0-(\kappa^3 c_7+\kappa^4 c_9+\cdots)]
-\frac{\kappa^2}{1+\kappa^2+\kappa^4}\mathcal{N}_5\tilde{c}_5. 
\end{equation}
By using this expression, we can eliminate $c_5$ from both of the right-hand sides of (\ref{c1}) and (\ref{c3}). 
Therefore, by repeating this procedure, the operators, $c_1,c_3,\ldots,c_{2N-1}$, can be 
expressed in terms of $\gamma_0,\tilde{c}_3,\tilde{c}_5,\ldots,\tilde{c}_{2N-1}$. 
Since $c_{2\ell}=\tilde{c}_{2\ell}$, the interaction potential $V$ can be expressed in terms of 
the operators, $\gamma_0,\tilde{c}_2,\tilde{c}_3,\tilde{c}_4,\ldots,\tilde{c}_{2N-2},\tilde{c}_{2N-1}$. 
Then, the resulting terms in the interaction decay by exponential law with the distance about these operators 
because the interaction $V$ which is written in terms of $c_1,c_2,\ldots,c_{2N-1}$ is assumed to be finite range. 
Further, {from} (\ref{etaRI}) and (\ref{eta}), one has 
\begin{equation}
\tilde{c}_i=\frac{1}{\sqrt{2}}\Bigl[\eta_i+\eta_i^\dagger -\sum_{j=2}^{2N-1}s(\tilde{A})_{i,j}(\eta_j-\eta_j^\dagger)\Bigr]
\end{equation}
for $i=2,3,\ldots,2N-1$. We recall $\hat{\gamma}_0=a_{\rm E}+a_{\rm E}^\dagger$. From these observations, 
one notices that the interaction $V$ can be written in terms of $a_{\rm E}, \eta_2,\eta_3,\ldots,\eta_{2N-1}$. 
In the following, we denote the interaction by 
\begin{equation}
\tilde{V}:={V}(a_{\rm E},\eta_2,\eta_3,\ldots,\eta_{2N-1}).
\end{equation}

The total Hamiltonian which we consider is given by 
\begin{equation}
\tilde{H}_g:=\tilde{H}_0+g\tilde{V},
\end{equation}
where $\tilde{H}_0$ is given by (\ref{tildeH0}). From the definitions (\ref{H0ctilde}) and (\ref{tildeH0p}), 
this right-hand side is written 
$$
\tilde{H}_0+g\tilde{V}=H_0+gV -\tilde{d}\tilde{A}\tilde{d}. 
$$
Clearly, the first and second terms in the right-hand side are those of the original Hamiltonian $H_g$, 
and the third term is the additional Hamiltonian whose spectrum has the gap above the unique ground state. 
Thus, it is enough to show that the spectrum of the Hamiltonian $\tilde{H}_g$ has 
a non-vanishing spectral gap above the two low-energy states. 

As mentioned at the beginning of this section, we assume that for a small $|g|$, 
there exists a spectral gap $\Delta\tilde{E}>0$ above the two-fold degenerate ground state which continuously 
connects with the spectral gap in the case of $g=0$.  
Both of the estimates of $|g|$ and $\Delta\tilde{E}$ will be obtained below.  

We write $|\tilde{0}\rangle$ for the ground state of $\tilde{H}_0$ of (\ref{tildeH0}) 
which satisfies $\eta_i|\tilde{0}\rangle =0$ for all $i=2,3,\ldots,2N-1$ and $a_{\rm E}|\tilde{0}\rangle=0$ 
for the operator $a_{\rm E}$ of the Majorana edge zero mode. 
Then, the other ground state is given by $a_{\rm E}^\dagger|\tilde{0}\rangle$.  
We also write $\tilde{P}_0(0)$ for the projection onto the sector of these two ground states. 
Under the above assumption on the spectral gap $\Delta\tilde{E}>0$ above the two low-energy states 
of $\tilde{H}_g$, there exists a unitary operator $\tilde{U}(g)$ such that \cite{Kato} 
\begin{equation}
\tilde{U}^\dagger(g)\tilde{P}_0(g)\tilde{U}(g)=\tilde{P}_0(0),
\end{equation}
where $\tilde{P}_0(g)$ is the projection onto the sector of the two low-energy states of $\tilde{H}_g$. 

Note that 
$$
\tilde{H}_g =[1-\tilde{P}_0(g)]\tilde{H}_g[1-\tilde{P}_0(g)]+\tilde{P}_0(g)\tilde{H}_g\tilde{P}_0(g). 
$$
Therefore, one has 
\begin{eqnarray*}
\tilde{H}(g)&:=&\tilde{U}^\dagger(g)\tilde{H}_g\tilde{U}(g)\nonumber\\
&=&[1-\tilde{P}_0(0)]\tilde{U}^\dagger(g)\tilde{H}_g\tilde{U}(g)[1-\tilde{P}_0(0)]
+\tilde{P}_0(0)\tilde{U}^\dagger (g)\tilde{H}_g\tilde{U}(g)\tilde{P}_0(0). 
\end{eqnarray*}
This implies that $\tilde{P}_0(0)$ is the projection onto the two low-energy states 
for the transformed Hamiltonian $\tilde{H}(g)$. In addition, the vector $|\tilde{0}\rangle$ is 
the eigenvector of $\tilde{H}(g)$. In fact, one has 
$$
\langle \tilde{0}|a_{\rm E}\tilde{U}^\dagger (g)\tilde{H}_g\tilde{U}(g)|\tilde{0}\rangle=0
$$
because $\tilde{U}^\dagger (g)\tilde{H}_g\tilde{U}(g)$ has even fermion parity by the construction \cite{BMNS} of 
the unitary operator $\tilde{U}(g)$. 
Thus, it is sufficient to show that 
there exists a non-vanishing spectral gap above the two low-energy states for this Hamiltonian $\tilde{H}(g)$. 

We write 
$$
\tilde{H}_0=\sum_{Z\subset\Lambda} \tilde{H}_{0,Z},
$$
where $\Lambda=\{1,2,\ldots,2N-1\}$, and $\tilde{H}_{0,Z}$ is the local Hamiltonian whose explicit form 
is given by 
$$
\tilde{H}_{0,Z}=2\eta_i^\dagger|\tilde{A}|_{i,j}\eta_j
$$
with $Z=\{i,j\}$ for $i,j=2,3,\ldots,2N-1$. Clearly, one has 
\begin{equation}
\tilde{H}_{0,Z}|\tilde{0}\rangle =0
\end{equation}
for all $Z$. Similarly, we can write the interaction $\tilde{V}$ in the sum of the local interactions $\tilde{V}_Z$, 
$$
\tilde{V}=\sum_{Z\subset\Lambda}\tilde{V}_Z, 
$$
where the support of $a_{\rm E}$ is given by the single site $\{1\}$. 
We write 
\begin{equation}
\label{tildeH0g}
\tilde{H}_0(g):=\tilde{U}^\dagger(g)\tilde{H}_0\tilde{U}(g)=\sum_{Z\subset\Lambda}\tilde{U}^\dagger(g)\tilde{H}_{0,Z}\tilde{U}(g)
\end{equation}
and 
\begin{equation}
\label{tildeVg}
\tilde{V}(g):=\tilde{U}^\dagger(g)\tilde{V}\tilde{U}(g)=\sum_{Z\subset\Lambda}\tilde{U}^\dagger(g)\tilde{V}_{Z}\tilde{U}(g).
\end{equation}
Obviously, 
\begin{equation}
\label{tildeHg}
\tilde{H}(g)=\tilde{H}_0(g)+g\tilde{V}(g).
\end{equation}

Following Hastings \cite{Hastings}, we introduce 
$$
\mathcal{R}_g(\cdots):=\int_{-\infty}^{+\infty} \tilde{\tau}_{g,t}(\cdots)w(t)dt, 
$$
where 
$$
\tilde{\tau}_{g,t}(\cdots):=\exp[it\tilde{H}(g)](\cdots)\exp[-it\tilde{H}(g)],
$$
and $w(t)$ is a function such that its Fourier transform $\hat{w}$ satisfies $\hat{w}(0)=1$ and $\hat{w}(\omega)=0$ for 
$|\omega|\ge\Delta\tilde{E}$. From the condition $\hat{w}(0)=1$, one has $\mathcal{R}_g(1)=1$ and 
\begin{equation}
\label{RtildeHg}
\mathcal{R}_g(\tilde{H}(g))=\tilde{H}(g).
\end{equation} 
Since the Hamiltonian $\tilde{H}(g)$ has the spectral gap $\Delta\tilde{E}>0$ above 
the sector of the two low-energy states from the assumption on the Hamiltonian $\tilde{H}_g$, 
the condition $\hat{w}(\omega)=0$ for $|\omega|\ge\Delta\tilde{E}$ yields  
\begin{equation}
\label{1-P0calAP0}
[1-\tilde{P}_0(0)]\mathcal{R}_g(\mathcal{A})\tilde{P}_0(0)=0
\end{equation}
for any operator $\mathcal{A}$. 
As a concrete function, we use the function $w(t)$ of the equation (2.1) in
Lemma~2.3 in \cite{BMNS}. From (\ref{tildeH0g}), (\ref{tildeVg}), (\ref{tildeHg}) and (\ref{RtildeHg}), we have 
\begin{equation}
\label{tildeHgdecomp}
\tilde{H}(g)=\sum_{Z\subset\Lambda}\mathcal{R}_g(\tilde{U}^\dagger(g)\tilde{H}_{0,Z}\tilde{U}(g))
+g \sum_{Z\subset\Lambda}\mathcal{R}_g(\tilde{U}^\dagger(g)\tilde{V}_Z\tilde{U}(g)). 
\end{equation}

Consider first the summand in the second sum in the right-hand side of (\ref{tildeHgdecomp}). 
Note that 
\begin{eqnarray*}
\mathcal{R}_g(\tilde{U}^\dagger(g)\tilde{V}_Z\tilde{U}(g))|\tilde{0}\rangle
&=&\tilde{P}_0(0)\mathcal{R}_g(\tilde{U}^\dagger(g)\tilde{V}_Z\tilde{U}(g))|\tilde{0}\rangle\\
&=&\int_{-\infty}^{+\infty}dt\; w(t)\tilde{P}_0(0)e^{it\tilde{H}(g)}\tilde{P}_0(0)
\tilde{U}^\dagger(g)\tilde{V}_Z\tilde{U}(g)|\tilde{0}\rangle e^{-it\tilde{E}_0(g)}\\
&=&\int_{-\infty}^{+\infty}dt\; w(t)|\tilde{0}\rangle\langle\tilde{0}|\tilde{U}^\dagger(g)\tilde{V}_Z\tilde{U}(g)|\tilde{0}\rangle\\
&=&|\tilde{0}\rangle\langle\tilde{0}|\tilde{U}^\dagger(g)\tilde{V}_Z\tilde{U}(g)|\tilde{0}\rangle,
\end{eqnarray*}
where $\tilde{E}_0(g)$ is the energy eigenvalue of $\tilde{H}(g)$ for the eigenvector $|\tilde{0}\rangle$, 
and we have used (\ref{1-P0calAP0}) and $\hat{w}(0)=1$; we have also used $\langle\tilde{0}|a_{\rm E}
\tilde{U}^\dagger(g)\tilde{V}_Z\tilde{U}(g)|\tilde{0}\rangle$=0, which holds because 
the operator $\tilde{U}^\dagger(g)\tilde{V}_Z\tilde{U}(g)$ has even fermion parity. 
This can be written 
\begin{equation}
\mathcal{R}_g(:\tilde{U}^\dagger(g)\tilde{V}_Z\tilde{U}(g):)|\tilde{0}\rangle=0,
\end{equation}
where $:(\cdots):$ is the subtraction of the expectation value with respect to the state $|\tilde{0}\rangle$ 
which is defined by 
\begin{equation}
:\mathcal{A}:=\mathcal{A}-\langle\tilde{0}|\mathcal{A}|\tilde{0}\rangle
\end{equation}
for an operator $\mathcal{A}$. We write 
\begin{equation}
\label{WZ1}
\tilde{\mathcal{W}}_Z^{(1)}(g):=g \mathcal{R}_g(:\tilde{U}^\dagger(g)\tilde{V}_Z\tilde{U}(g):).
\end{equation}

Next consider the summand in the first sum in the right-hand side of (\ref{tildeHgdecomp}). 
Note that 
\begin{eqnarray}
\label{diffUH0ZU-H0Z}
\tilde{U}^\dagger(g)\tilde{H}_{0,Z}\tilde{U}(g)-\tilde{H}_{0,Z}
&=& \int_0^g dg' \frac{d}{dg'} \tilde{U}^\dagger(g')\tilde{H}_{0,Z}\tilde{U}(g')\nonumber\\
&=&\int_0^g dg' \tilde{U}^\dagger(g')i[\tilde{H}_{0,Z},\tilde{D}(g')]\tilde{U}(g'), 
\end{eqnarray}
where we have used \cite{BMNS} 
$$
\frac{d}{dg}\tilde{U}(g)=i\tilde{D}(g)\tilde{U}(g)
$$
with the self-adjoint operator $\tilde{D}(g)$. Therefore, we have 
\begin{equation}
\label{calRgUH0ZU}
\mathcal{R}_g(\tilde{U}^\dagger(g)\tilde{H}_{0,Z}\tilde{U}(g))=\mathcal{R}_g(\tilde{H}_{0,Z})
+\int_0^g dg' \mathcal{R}_g(\tilde{U}^\dagger(g')i[\tilde{H}_{0,Z},\tilde{D}(g')]\tilde{U}(g')).
\end{equation}
The second term in the right-hand side satisfies 
$$
\int_0^g dg' \mathcal{R}_g(:\tilde{U}^\dagger(g')i[\tilde{H}_{0,Z},\tilde{D}(g')]\tilde{U}(g'):)|\tilde{0}\rangle=0
$$
in the same way as in the above. Therefore, we write 
\begin{equation}
\label{WZ2}
\tilde{\mathcal{W}}_Z^{(2)}(g):=
\int_0^g dg' \mathcal{R}_g(:\tilde{U}^\dagger(g')i[\tilde{H}_{0,Z},\tilde{D}(g')]\tilde{U}(g'):).
\end{equation}

In order to deal with the first term $\mathcal{R}_g(\tilde{H}_{0,Z})$ in the right-hand side of (\ref{calRgUH0ZU}), 
we note that 
\begin{eqnarray}
\label{eHgH0ZeHg-H0Z}
e^{it\tilde{H}(g)}\tilde{H}_{0,Z}e^{-it\tilde{H}(g)}-\tilde{H}_{0,Z}
&=&\int_0^t dt'\frac{d}{dt'}e^{it'\tilde{H}(g)}\tilde{H}_{0,Z}e^{-it'\tilde{H}(g)}\nonumber\\
&=&\int_0^t dt'\; e^{it'\tilde{H}(g)}i[\tilde{H}(g),\tilde{H}_{0,Z}]e^{-it'\tilde{H}(g)}.
\end{eqnarray}
On the other hand, the above relation (\ref{diffUH0ZU-H0Z}) yields 
\begin{eqnarray*}
\tilde{H}(g)&=&\tilde{U}^\dagger(g)\tilde{H}_0\tilde{U}(g)+g\tilde{U}^\dagger(g)\tilde{V}\tilde{U}(g)\\
&=&\tilde{H}_0+\int_0^g dg'\; \tilde{U}^\dagger(g')i[\tilde{H}_0,\tilde{D}(g')]\tilde{U}(g')
+g\tilde{U}^\dagger(g)\tilde{V}\tilde{U}(g).
\end{eqnarray*}
Substituting this into the right-hand side of (\ref{eHgH0ZeHg-H0Z}), we have 
\begin{eqnarray}
\label{eHgHoZeHg}
e^{it\tilde{H}(g)}\tilde{H}_{0,Z}e^{-it\tilde{H}(g)}&=&\tilde{H}_{0,Z}+
\int_0^t dt'\; e^{it'\tilde{H}(g)}i[\tilde{H}_0,\tilde{H}_{0,Z}]e^{-it'\tilde{H}(g)}+\tilde{\mathcal{M}}_{t,Z}(g), 
\end{eqnarray}
where 
\begin{equation}
\tilde{\mathcal{M}}_{t,Z}(g):=\int_0^t dt'\; \tilde{\tau}_{g,t'}
\left(ig[\tilde{U}^\dagger(g)\tilde{V}\tilde{U}(g),\tilde{H}_{0,Z}] 
+ \int_0^g dg'\; i[\tilde{U}^\dagger(g')i[\tilde{H}_0,\tilde{D}(g')]\tilde{U}(g'),\tilde{H}_{0,Z}]\right).
\end{equation}
{From} the expression of (\ref{eHgHoZeHg}), one notices that 
\begin{equation}
\tilde{\mathcal{M}}_{t,Z}(g)|\tilde{0}\rangle=0.  
\end{equation}
Actually, $|\tilde{0}\rangle$ is the eigenvector of $\tilde{H}(g)$, and 
$\tilde{H}_{0,Z}|\tilde{0}\rangle=0$. In addition, we obtain 
\begin{equation}
\label{sumRgH0Z}
\sum_{Z\subset\Lambda}\mathcal{R}_g(\tilde{H}_{0,Z})=\tilde{H}_0
+\sum_{Z\subset\Lambda}\int_{-\infty}^{+\infty}dt\; w(t) \tilde{\mathcal{M}}_{t,Z}(g).
\end{equation} 
We write 
\begin{equation}
\label{WZ3}
\tilde{\mathcal{W}}_Z^{(3)}(g):=\int_{-\infty}^{+\infty}dt\; w(t) \tilde{\mathcal{M}}_{t,Z}(g).
\end{equation}

Consequently, from (\ref{tildeHgdecomp}), (\ref{WZ1}), (\ref{calRgUH0ZU}), (\ref{WZ2}), (\ref{sumRgH0Z}) and (\ref{WZ3}), 
we obtain the desired expression of the Hamiltonian $\tilde{H}(g)$, 
\begin{equation}
\tilde{H}(g)=\tilde{H}_0+\tilde{\mathcal{W}}(g),
\end{equation}
with 
\begin{equation}
\tilde{\mathcal{W}}(g):=\sum_{i=1}^3 \sum_{Z\subset\Lambda}\tilde{\mathcal{W}}_Z^{(i)}(g), 
\end{equation}
where we have dropped the constant term. The operators $\tilde{\mathcal{W}}_Z^{(i)}(g)$ satisfy 
\begin{equation}
\tilde{\mathcal{W}}_Z^{(i)}(g)\rightarrow 0 \quad \mbox{as \ \ } g\rightarrow 0, 
\end{equation}
and
\begin{equation}
\tilde{\mathcal{W}}_Z^{(i)}(g)|\tilde{0}\rangle=0
\end{equation}
for all $Z$ and $i=1,2,3$. Therefore, in the same way as in \cite{Koma}, we obtain 
\begin{equation}
[\tilde{\mathcal{W}}(g)]^2\le (\tilde{g})^2 (\tilde{\mathfrak{N}})^2,
\end{equation}
where $\tilde{g}$ is a positive constant which satisfies $\tilde{g}\rightarrow 0$ as $g\rightarrow 0$, 
and 
\begin{equation}
\tilde{\mathfrak{N}}:=a_{\rm E}^\dagger a_{\rm E}+\sum_{i=2}^{2N-1}\eta_i^\dagger \eta_i. 
\end{equation}
By using the bound (\ref{lowerboundH0}), we have 
\begin{eqnarray}
\label{tildecalWg2Lbound}
[\tilde{\mathcal{W}}(g)]^2&\le& (\tilde{g})^2 (\tilde{\mathfrak{N}})^2\nonumber\\
&=&(\tilde{g})^2\Bigl(a_{\rm E}^\dagger a_{\rm E}+\sum_i \eta_i^\dagger\eta_i\Bigr)^2\nonumber\\
&\le&(\tilde{g})^2\Bigl(1+\frac{1}{\Delta E_0}\tilde{H}_0\Bigr)^2\nonumber\\
&\le&2(\tilde{g})^2\Bigl[1+\frac{1}{(\Delta E_0)^2}(\tilde{H}_0)^2\Bigr].
\end{eqnarray}
 
By the construction of the Hamiltonian $\tilde{H}(g)=\tilde{H}_0+\tilde{\mathcal{W}}(g)$, one has 
$$
\langle \tilde{0}|\tilde{H}(g)|\tilde{0}\rangle=\langle\tilde{0}|\tilde{H}_0|\tilde{0}\rangle 
+\langle\tilde{0}|\tilde{\mathcal{W}}(g)|\tilde{0}\rangle=0.
$$
This implies that the ground state energy $\tilde{E}_0(g)$ of $\tilde{H}(g)$ satisfies 
$\tilde{E}_0(g)\le 0$. 

Let $\tilde{\Psi}_1$ be an excited state of $\tilde{H}(g)$ such that the energy eigenvalue $\tilde{E}_1(g)$ 
does not connect continuously with the ground-state energy zero of the unperturbed Hamiltonian $\tilde{H}_0$ 
when varying the coupling constant $g$ from non-zero to zero. From $\tilde{H}(g)\tilde{\Psi}_1=\tilde{E}_1(g)\tilde{\Psi}_1$ 
with the normalization $\Vert\tilde{\Psi}_1\Vert=1$, one has  
\begin{eqnarray}
[\tilde{E}_1(g)]^2&=&\langle \tilde{\Psi}_1,[\tilde{H}(g)]^2\tilde{\Psi}_1\rangle\nonumber\\
&=&\langle\tilde{\Psi}_1,(\tilde{H}_0)^2\tilde{\Psi}_1\rangle 
+\langle\tilde{\Psi}_1,[\tilde{\mathcal{W}}(g)]^2\tilde{\Psi}_1\rangle
+\langle\tilde{\Psi}_1,\tilde{H}_0\tilde{\mathcal{W}}(g)\tilde{\Psi}_1\rangle
+\langle\tilde{\Psi}_1,\tilde{\mathcal{W}}(g)\tilde{H}_0\tilde{\Psi}_1\rangle\nonumber\\
&\ge&\langle\tilde{\Psi}_1,(\tilde{H}_0)^2\tilde{\Psi}_1\rangle
+\langle\tilde{\Psi}_1,\tilde{H}_0\tilde{\mathcal{W}}(g)\tilde{\Psi}_1\rangle
+\langle\tilde{\Psi}_1,\tilde{\mathcal{W}}(g)\tilde{H}_0\tilde{\Psi}_1\rangle. 
\end{eqnarray}
The second and third terms in the right-hand side can be estimated as follows: 
\begin{eqnarray}
\bigl|{\rm Re}\langle \tilde{\Psi}_1,\tilde{H}_0\tilde{\mathcal{W}}(g)\tilde{\Psi}_1\rangle\bigr|
&\le& \sqrt{\langle\tilde{\Psi}_1,(\tilde{H}_0)^2\tilde{\Psi}_1\rangle 
\langle\tilde{\Psi}_1,[\tilde{\mathcal{W}}(g)]^2\tilde{\Psi}_1\rangle   }\nonumber\\
&\le&\frac{\sqrt{2}\tilde{g}}{\Delta E_0}\sqrt{\langle\tilde{\Psi}_1,(\tilde{H}_0)^2\tilde{\Psi}_1\rangle 
\langle\tilde{\Psi}_1,[(\Delta E_0)^2 +(\tilde{H}_0)^2]\tilde{\Psi}_1\rangle   }\nonumber\\
&\le&\sqrt{2}\tilde{g}\Delta E_0 +\frac{\sqrt{2}\tilde{g}}{\Delta E_0}\langle\tilde{\Psi}_1,(\tilde{H}_0)^2\tilde{\Psi}_1\rangle,
\end{eqnarray}
where we have used the bound (\ref{tildecalWg2Lbound}). 
Substituting this into the above right-hand side, we have 
$$
[\tilde{E}_1(g)]^2+2\sqrt{2}\tilde{g}\Delta E_0
\ge \left[1-\frac{2\sqrt{2}\tilde{g}}{\Delta E_0}\right]\langle\tilde{\Psi}_1,(\tilde{H}_0)^2\tilde{\Psi}_1\rangle. 
$$
Further, using the bound (\ref{tildecalWg2Lbound}) again, we obtain 
\begin{eqnarray}
{2(\tilde{g})^2}\frac{[\tilde{E}_1(g)]^2+2\sqrt{2}\tilde{g}\Delta E_0}{{(\Delta E_0)^2}-{2\sqrt{2}\tilde{g}}{\Delta E_0}}
+2(\tilde{g})^2
&\ge&\langle\tilde{\Psi}_1,[\tilde{\mathcal{W}}(g)]^2\tilde{\Psi}_1\rangle \nonumber\\
&=&\langle\tilde{\Psi}_1,[\tilde{H}(g)-\tilde{H}_0]^2\tilde{\Psi}_1\rangle \nonumber\\
&=&\langle\tilde{\Psi}_1,[\tilde{E}_1(g)-\tilde{H}_0]^2\tilde{\Psi}_1\rangle. 
\end{eqnarray}
Since we can assume $\tilde{E}_1(g)\le \Delta E_0$, we have the bound, 
\begin{equation}
\label{diffE1H0bound}
\langle\tilde{\Psi}_1,[\tilde{E}_1(g)-\tilde{H}_0]^2\tilde{\Psi}_1\rangle
\le \frac{4\Delta E_0}{\Delta E_0-2\sqrt{2}\tilde{g}}(\tilde{g})^2. 
\end{equation}
When $\tilde{E}_1(g)$ satisfies $\Delta E_0\ge \tilde{E}_1(g)>\Delta E_0/2$ for a small $|g|$, one has 
$$
\frac{1}{2} \Delta E_0 > \sqrt{\langle\tilde{\Psi}_1,[\tilde{E}_1(g)-\tilde{H}_0]^2\tilde{\Psi}_1\rangle} 
$$
{from} the spectrum of $\tilde{H}_0$. Therefore, {from} the above bound (\ref{diffE1H0bound}), we obtain 
\begin{equation}
\label{tildeE1gbound}
\frac{1}{2}\Delta E_0 <\Delta E_0-\left(\frac{4\Delta E_0}{\Delta E_0-2\sqrt{2}\tilde{g}}\right)^{1/2} \tilde{g}
\le \tilde{E}_1(g)\le \Delta E_0
\end{equation}
for a small $|g|$. This implies that there exists a small positive constant $g_{\rm max}$ 
such that the bound (\ref{tildeE1gbound}) can be realized for any coupling constant $g$ 
which satisfies $|g|\le g_{\rm max}$.

\Section{Proof of the existence of $\gamma$ satisfying $[H,\gamma]=0$}
\label{existencegamma} 

In this appendix, we give a proof of the existence of a Majorana operator $\gamma$ 
which commutes with the Hamiltonian $H$, i.e., $[H,\gamma]=0$, 
following the paper \cite{GC} by Goldstein and Chamon. 
The Hamiltonian $H$ has even fermion parity and 
consists of the operators, $c_1,c_2,\ldots,c_{2N-1}$, whose number is odd $2N-1$. 

Let us consider all the monomials which have odd fermion parity, and 
consist of the operators, $c_1,c_2,\ldots,c_{2N-1}$.    
There are in total $\sum_{m=1}^N {2N-1 \choose 2m-1}=2^{2N-2}$ such operators. 
We denote them by $C_a$ with indices $a$. We can write the operators $C_a$ in the form, 
\begin{equation}
C_a:=(i)^{n_a(n_a-1)/2}c_{i_1(a)}c_{i_2(a)}\cdots c_{i_{n_a}(a)},
\end{equation}
where $n_a$ is the number of $c$-Majorana fermions in the product, 
and for each monomial with index $a$, 
$I(a):=\{i_1(a),i_2(a),\ldots,i_{n_a}(a)\}$ is the list of the sites which satisfy conditions, 
$i_j(a)\in\{1,2,\cdots,2N-1\}$ for $j\in\{1,2,\ldots,n_a\}$ and $i_k(a)<i_\ell(a)$ for $k<\ell$. 
Of course, $n_a$ is odd and satisfies $n_a\le 2N-1$. Then, one has 
$$
C_a^\dagger=C_a \quad \mbox{and}\quad (C_a)^2=1.
$$
Clearly, the operator $\gamma$ is written as 
\begin{equation}
\gamma=\sum_a \alpha_a C_a
\end{equation}
with coefficients $\alpha_a$. 

The product of two $C_{a_1}$ and $C_{a_2}$ gives 
$$
C_{a_1}C_{a_2}=(i)^{s(a_1,a_2)}C_{a_3}
$$
with some phase factor, where the set $I(a_3)$ of the sites in $C_{a_3}$  
satisfies $I(a_3)=I(a_1)\cup I(a_2)\backslash I(a_1)\cap I(a_2)$. 
We introduce the usual trace inner product for two operators, $A$ and $B$: 
$$
(A,B):=\frac{1}{2^{2N}}{\rm Tr}A^\dagger B.
$$
Then, one has 
\begin{equation}
(C_a,C_b)=\delta_{a,b}
\end{equation}
for all $C_a$ and $C_b$. Actually, 
$$
{\rm Tr}\; c_{j_1}c_{j_2}\cdots c_{j_n} =0 
$$
for any monomial $c_{j_1}c_{j_2}\cdots c_{j_n}$ with $n\ge1$. In the case of $n={\rm odd}$, this is true 
because of the odd fermion parity. When $n={\rm even}$, one has 
$$
{\rm Tr}\; c_{j_1}c_{j_2}\cdots c_{j_n}={\rm Tr}\; c_{j_2}\cdots c_{j_n}c_{j_1}
=(-1)^{n-1}{\rm Tr}\; c_{j_1}c_{j_2}\cdots c_{j_n}=-{\rm Tr}\; c_{j_1}c_{j_2}\cdots c_{j_n}
$$
by using the cyclic property of trace and the commutation relations. This implies the vanishing of the trace. 

Let us consider 
\begin{equation}
H_{a,b}:=(C_a,[H,C_b])
\end{equation}
for the Hamiltonian $H$ with even fermion parity. This is a $2^{2N-2}\times 2^{2N-2}$ matrix.  
Further, this is pure imaginary and antisymmetric. Actually, one has 
$$
\overline{H_{a,b}}=\overline{(C_a,[H,C_b])}=([H,C_b],C_a)
=2^{-2N}{\rm Tr}(-[H,C_b])C_a=-2^{-2N}{\rm Tr}\; C_a[H,C_b]=-H_{a,b}.
$$
Thus, $H_{a,b}$ is pure imaginary. Further, the cyclic property of trace yields  
$$
{\rm Tr}\; C_a[H,C_b]={\rm Tr}\; (C_aHC_b-C_aC_bH)={\rm Tr}\;C_bC_aH-{\rm Tr}\; C_bHC_a.
=-{\rm Tr}\; C_b[H,C_a]
$$
This implies $H_{a,b}=-H_{b,a}$, i.e., $H_{a,b}$ is antisymmetric. 

One can easily show $[H,c_1c_2\cdots c_{2N-1}]=0$. Therefore, the matrix $H_{a,b}$ has the corresponding 
zero eigenvalue. Since the matrix $H_{a,b}$ is pure imaginary and antisymmetric with the even dimension $2^{2N-2}$, 
there must exist at least one more zero eigenvalue. Namely, there exists an eigenvector $\alpha_b$ such that 
it satisfies the eigenvalue equation, 
\begin{equation}
0=\sum_b H_{a,b}\alpha_b=
\sum_b (C_a,[H,C_b])\alpha_b=(C_a,[H,\gamma]),
\end{equation}
where we have written 
\begin{equation}
\gamma=\sum_b \alpha_b C_b.
\end{equation}
The above equation must hold for any $C_a$. Therefore, one obtains $[H,\gamma]=0$, 
and $\gamma\ne c_1c_2\cdots c_{2N-1}$.

\Section{Locality of Majorana edge zero mode}
\label{LocalMajoranaedge}

In this appendix, we prove the locality of the operator $U(g)\hat{\gamma}_0 U(g)^\dagger$, 
which creates the Majorana edge zero mode above the ground state. 
{From} (\ref{gamma0}) and (\ref{hatgamma0}), it is sufficient to prove the locality of 
$$
U(g)c_{2\ell-1} U(g)^\dagger
$$
for a not large $\ell\in\{1,2,\ldots,N\}$. For this purpose, we will use a local approximation \cite{BHV,BMNS,NSY}. 

To begin with, we recall that the Hamiltonian $H_g$ does not contain the operator $c_{2N}$ at the right edge. 
This is crucial for the following argument. Note that
$$
(c_{2m-1}c_{2m})c_{2m-1}(c_{2m}c_{2m-1})=-c_{2m-1} \quad \mbox{for \ } m=1,2,\ldots, N,
$$
$$
(c_{2m-1}c_{2m})c_{2m}(c_{2m}c_{2m-1})=-c_{2m} \quad \mbox{for \ } m=1,2,\ldots, N-1,
$$
$$
(c_{2m-1}c_{2N})c_{2m-1}(c_{2N}c_{2m-1})=-c_{2m-1}\quad \mbox{for \ } m=1,2,\ldots, N-1,
$$
and 
$$
(c_{2m-1}c_{2N})c_{2m-1}c_{2m}(c_{2N}c_{2m-1})=-c_{2m-1}c_{2m} \quad \mbox{for \ } m=1,2,\ldots, N-1.
$$
Namely, these unitary transformations change the sign of the operators. 
Having these relations in mind, we introduce a unitary transformation,   
$$
\mathcal{U}_{\ge m}(\sigma_m,\sigma_{m+1},\ldots, \sigma_N):=\mathcal{U}_m(\sigma_m)\mathcal{U}_{m+1}(\sigma_{m+1})
\cdots \mathcal{U}_{N}(\sigma_{N}), 
$$
for $m=1,2,\ldots,N$, where 
$$
\mathcal{U}_m(\sigma_m):=
\begin{cases}
1, &  \sigma_m=1;\\
c_{2m}c_{2m-1}, &  \sigma_m =2 ;\\
c_{2N}c_{2m-1}, &  \sigma_m =3;\\
c_{2m}c_{2m-1}c_{2N}c_{2m-1}, & \sigma_m=4, 
\end{cases}
$$
for $m=1,2,\ldots,N-1$, and 
$$
\mathcal{U}_N(\sigma_N):=
\begin{cases}
1, & \sigma_N=1;\\
c_{2N}c_{2N-1}, & \sigma_N=2.
\end{cases} 
$$
By definition, the operator $\mathcal{U}_{\ge m}(\sigma_m,\sigma_{m+1},\ldots, \sigma_N)$ has even fermion parity. 
For any operator $A$, we define 
\begin{equation}
\label{Pi}
\Pi_{\ge 2m-1}(A):=\frac{1}{2}\left(\frac{1}{4}\right)^{N-m}\sum_{\sigma_N=1,2}\;\sum_{\sigma_m,\sigma_{m+1},\ldots,\sigma_{N-1}}
\mathcal{U}_{\ge m}^\dagger(\sigma_m,\ldots, \sigma_N)A\mathcal{U}_{\ge m}(\sigma_m,\ldots, \sigma_N).
\end{equation}
This is the average over the unitary transformations. 
When an operator $A$ is written in a product form $A=A'A''$ which satisfies ${\rm supp}A'\subset\{1,2,\ldots,2m-2\}$ 
and ${\rm supp}A''\subset\{2m-1,2m,\ldots,2N-1\}$, one has $\Pi_{\ge 2m-1}(A)=A'\Pi_{\ge 2m-1}(A'')$.   
If the operator $A\ne 1$ satisfies the condition ${\rm supp}A\subset\{2m-1,2m,\ldots,2N-1\}$, 
then $\Pi_{\ge 2m-1}(A)=0$. Of course, if $A=1$, then $\Pi_{\ge 2m-1}(A)=1$. 

For the operator $U(g)c_{2\ell-1} U(g)^\dagger$, the local approximation is given by 
$$
\Pi_{\ge 2m-1}(U(g)c_{2\ell-1} U(g)^\dagger).
$$
{From} the definition(\ref{Pi}) of $\Pi_{\ge 2m-1}(\cdots)$, the difference between the two operators can be written 
\begin{eqnarray}
\label{diffUgcUg}
& &U(g)c_{2\ell-1} U(g)^\dagger-\Pi_{\ge 2m-1}(U(g)c_{2\ell-1} U(g)^\dagger)\nonumber\\
&=&\frac{1}{2}\left(\frac{1}{4}\right)^{N-m}\sum_{\sigma_m,\ldots,\sigma_N}
\mathcal{U}_{\ge m}^\dagger(\sigma_m,\ldots, \sigma_N)
\left[\mathcal{U}_{\ge m}(\sigma_m,\ldots, \sigma_N),U(g)c_{2\ell-1} U(g)^\dagger\right]
\end{eqnarray}
by using the commutator. In order to evaluate the commutator, we recall the following facts \cite{BMNS}: 
The unitary operator $U(g)$ is the solution of the Schr\"odinger-type equation, 
\begin{equation}
-i\frac{d}{dg}U(g)=D(g)U(g), 
\end{equation}
with the initial condition $U(0)=1$, where the self-adjoint operator $D(g)$ is written 
\begin{equation}
\label{Dg}
D(g)=\int_{-\infty}^{+\infty}dt\; W(t)\; e^{it H_g} V e^{-itH_g}
\end{equation}
with a certain real-valued function $W(t)$ which shows subexponential decay \cite{BMNS}.   
The operator $D(g)$ is correspond to the Hamiltonian of the Schr\"odinger-type equation. 
The range of the corresponding interaction of $D(g)$ is determined by the interaction Hamiltonian $V$ 
of the present Hamiltonian $H_g$ as seen in the right-hand side of $(\ref{Dg})$. 
When $V$ is finite-range or exponentially decaying interaction, $D(g)$ shows  
subexponentially decaying interaction \cite{BMNS}. 
Under the assumption on the range of the interaction $V$,  
the following Lieb-Robinson bound holds \cite{BMNS,Koma}:
\begin{equation}
\label{LRbound}
\Vert [U(g)AU(g)^\dagger,B]\Vert \le \Vert A\Vert \Vert B\Vert \sum_{i\in{\rm supp}A, j\in{\rm supp}B}F(|i-j|)
\end{equation}
for any operator $A$ with odd fermion parity, and any operator $B$ with even fermion parity,  
where $F$ is a subexponentially decaying function. Here, it is crucial that at least 
one of the operators, $A$ and $B$, must have even fermion parity \cite{HK,NSY,Koma}. 
Actually, if the two operators, $A$ and $B$, anticommute with each other for large distance, 
then the commutator in the Lieb-Robinson bound must be replaced by the anticommutator \cite{HK}. 
Combining the Lieb-Robinson bound (\ref{LRbound}) with (\ref{diffUgcUg}), we have 
\begin{eqnarray}
\label{diffUgellUgPibound}
& &\Vert U(g)c_{2\ell-1} U(g)^\dagger-\Pi_{\ge 2m-1}(U(g)c_{2\ell-1} U(g)^\dagger)\Vert \nonumber\\
&\le&\frac{1}{2}\left(\frac{1}{4}\right)^{N-m}\sum_{\sigma_m,\ldots,\sigma_N}
\left\Vert\left[\mathcal{U}_{\ge m}(\sigma_m,\ldots, \sigma_N),U(g)c_{2\ell-1} U(g)^\dagger\right]\right\Vert\nonumber\\
&\le& \sum_{n\ge 2m-1} F(|2\ell-1-n|)\le \tilde{F}(|\ell-m|) 
\end{eqnarray}
for $\ell\le m$, where we have used that the operator $\mathcal{U}_{\ge m}(\sigma_m,\ldots, \sigma_N)$ 
has even fermion parity as mentioned above, and $\tilde{F}$ is a subexponentially decaying function. 
This implies the locality of $U(g)c_{2\ell-1} U(g)^\dagger$. In fact, we have the local decomposition, \cite{BMNS}
\begin{equation}
U(g)c_{2\ell-1}U(g)^\dagger=\sum_{m=1}^\infty \Delta_m(c_{2\ell-1}),
\end{equation}
where 
$$
\Delta_1(c_{2\ell-1})=\Pi_{\ge 1}(U(g)c_{2\ell-1}U(g)^\dagger),
$$ 
and 
$$
\Delta_m(c_{2\ell-1})=\Pi_{\ge 2m-1}(U(g)c_{2\ell-1}U(g)^\dagger)-\Pi_{\ge 2m-3}(U(g)c_{2\ell-1}U(g)^\dagger)
$$
for $m\ge 2$. From the above bound (\ref{diffUgellUgPibound}), one has 
\begin{eqnarray*}
\Vert\Delta_m(c_{2\ell-1})\Vert
&=&\Vert\Pi_{\ge 2m-1}(U(g)c_{2\ell-1}U(g)^\dagger)-\Pi_{\ge 2m-3}(U(g)c_{2\ell-1}U(g)^\dagger)\Vert\\
&\le& \Vert\Pi_{\ge 2m-1}(U(g)c_{2\ell-1}U(g)^\dagger)-U(g)c_{2\ell-1}U(g)^\dagger\Vert\\
&+&\Vert U(g)c_{2\ell-1}U(g)^\dagger -\Pi_{\ge 2m-3}(U(g)c_{2\ell-1}U(g)^\dagger)\Vert\\
&\le&\tilde{F}(|\ell-m|)+\tilde{F}(|\ell-m+1|)
\end{eqnarray*}
for a large $m$. This implies a subexponentially decay. 


\end{document}